\documentclass[%
 aip,
 jap,
 amsmath,amssymb,
 reprint,%
]{revtex4-1}

\usepackage{graphicx}% Include figure files
\usepackage{dcolumn}% Align table columns on decimal point
\usepackage{bm}% bold math
\usepackage[utf8]{inputenc}
\usepackage[T1]{fontenc}
\usepackage{mathptmx}
\usepackage{upgreek}
\usepackage{gensymb}

\begin{document}

\preprint{AIP/123-QED}

\title[Directing Near-Infrared Photon Transport with Core@Shell Particles]{Directing Near-Infrared Photon Transport with Core@Shell Particles}

\author{Kevin M. Conley}
\affiliation{Department of Applied Physics, QTF Centre of Excellence, Aalto University School of Science, P.O. Box 11100, FI-00076 Aalto, Finland}
\author{Vaibhav Thakore}
\affiliation{Department of Applied Mathematics, The University of Western Ontario, London, Ontario N6A 5B7, Canada}
\affiliation{Center for Advanced Materials and Biomaterials Research, The University of Western Ontario, London, Ontario N6A 3K7, Canada}
\author{Fahime Seyedheydari}
\affiliation{Department of Applied Physics, QTF Centre of Excellence, Aalto University School of Science, P.O. Box 11100, FI-00076 Aalto, Finland}
\author{Mikko Karttunen}
\affiliation{Department of Applied Mathematics, The University of Western Ontario, London, Ontario N6A 5B7, Canada}
\affiliation{Center for Advanced Materials and Biomaterials Research, The University of Western Ontario, London, Ontario N6A 3K7, Canada}
\affiliation{Department of Chemistry, The University of Western Ontario, London, Ontario N6A 5B7, Canada}
\author{Tapio Ala-Nissila}
%\email{tapio.ala-nissila@aalto.fi}
%\phone{+358 40 5412983}
\affiliation{Department of Applied Physics, QTF Centre of Excellence, Aalto University School of Science, P.O. Box 11100, FI-00076 Aalto, Finland}
\affiliation{Interdisciplinary Centre for Mathematical Modelling, Department of Mathematical Sciences, Loughborough University, Loughborough, Leicestershire LE11 3TU, United Kingdom}

\date{\today}

\begin{abstract}
Directing the propagation of near-infrared radiation is a major concern in improving the efficiency of solar cells and thermal insulators. A facile approach to scatter light in the near-infrared region without excessive heating is to embed compact layers with semiconductor particles. The directional scattering by semiconductor@oxide (core@shell) spherical particles (containing Si, InP, TiO$_2$, SiO$_2$, or ZrO$_2$) with a total radius varying from 0.1 to 4.0~$\upmu$m and in an insulating medium at low volume fraction is investigated using Lorenz-Mie theory and multiscale modelling. The optical response of each layers is calculated under irradiation by the sun or a blackbody emitter at 1180~K. Reflectance efficiency factors of up to 83.7\% and 63.9\% are achieved for near-infrared solar and blackbody radiation in 200~$\upmu$m thick compact layers with only 1\% volume fraction of bare Si particles with a radius of 0.23~$\upmu$m and 0.50~$\upmu$m, respectively. The maximum solar and blackbody efficiency factors of layers containing InP particles was slightly less (80.2\% and 60.7\% for bare particles with a radius of 0.25~$\upmu$m and 0.60~$\upmu$m, respectively). The addition of an oxide coating modifies the surrounding dielectric environment, which improves the solar reflectance efficiency factor to over 90\% provided it matches the scattering mode energies with the incident spectral density. The layers are spectrally-sensitive and can be applied as a back or front reflector for solar devices, high temperature thermal insulators, and optical filters in Gradient Heat Flux Sensors for fire safety applications. 
\end{abstract}

\maketitle

\section{Introduction}\label{intro}
Controlled propagation of near-infrared (near-IR) electromagnetic waves has been demonstrated in diverse applications such as optical and biological sensors,~\cite{kabashin2009plasmonic,matsui2013oxide} photocatalysis for solar energy storage,~\cite{preston2008preparation} solar glazing,~\cite{sousa2017hybrid} and thermal energy management.~\cite{jonsson2017solar} These strategies use the large scattering cross-sections of the particles to manipulate the direction of light propagation. The near-IR region is the primary radiative power of flame and blackbody emitters and accounts for 52\% of the sun's irradiance power.~\cite{sousa2017hybrid}

The efficiency of high performance ultra thin solar cells is limited due to their poor absorption of near-IR radiation.~\cite{atwater2010plasmonics} One approach to mitigate the decreased physical thickness is to increase the optical thickness using directional scattering of light by particles embedded in layers.~\cite{atwater2010plasmonics} Light trapping at the rear or within dye sensitized solar cells has been simulated using Monte Carlo methods.~\cite{akimov2010resonant,galvez2012effect,guo2013optimization,sasanpour2014theoretical} It was shown that guiding or trapping light inside the device with a back-reflector layer improves the efficiency.~\cite{siebentritt2017chalcopyrite,yin2015integration,van2015light} 

A compact layer placed at the front of a device, such as a Gradient Heat Flux Sensor (GHFS), could regulate the transmittance of near-IR radiation into the device. GHFSs, based on anisotropic thermoelements, provide quantitative information about heat transport within confined spaces with fast response times ($10^{-9}$ to $10^{-8}$ s).~\cite{mityakov2012gradient} The transmittance can be enhanced using layers containing forward scattering particles~\cite{derkacs2006improved,matheu2008metal} or selectively limited using a reflective layer. Near-IR sources could be distinguished by selecting a wavelength range based on the characteristics of the spectral density, such as the CO$_2$ emission peak of an n-heptane pool fire.~\cite{bordbar2019numerical}

The dielectric response of homogeneous and core@shell spheres in an external electromagnetic field have closed-form analytic solutions.~\cite{bohren2008absorption,kerker2013scattering,tzarouchis2018light} Directional scattering has been demonstrated for metal@oxide (core@shell) spherical nanoparticles in the visible region.~\cite{ruffino2014simulations,pustovalov2015influence} The addition of a shell can enhance the radiative heat transfer between metal@oxide particles~\cite{nikbakht2018radiative} or improve the sensitivity of the mode frequency to changes in the refractive index of the medium.~\cite{zhang2018double} 

Conventional metals often require shape anisotropy to extend to the near-IR,~\cite{preston2008preparation,wallace2019advancements} Further, the effectiveness of metallic particles is limited by large Ohmic dissipation at high temperatures.~\cite{naik2013alternative} By contrast, low-bandgap semiconductor particles exhibit large scattering in the near-IR without resorting to internal structuring or non-spherical shapes.~\cite{tang2017plasmonically} Unlike metals, the scattering performance in semiconductor particles is maintained at elevated temperatures and the particles do not suffer from excessive heating.~\cite{thakore2019thermoplasmonic} The amenability of the semiconductor particle shape and orientation offers a straightforward method to generate reflective layers in the near-IR region. 

In this article, the optical properties of compact layers embedded with core@shell semiconductor particles (containing Si, InP, TiO$_2$, SiO$_2$, or ZrO$_2$) with a total radius varying from 0.1 to 4.0~$\upmu$m are investigated and characterized by their reflectance under irradiation from two near-IR sources (solar and blackbody emission). The dielectric response and directional scattering of core@shell spheres embedded in an insulating medium at low volume fraction (1\%) is calculated using Lorenz-Mie theory and a Monte Carlo method described in Section~\ref{methods}. The optimal bare particle radii to maximize the reflectance efficiency of layers under near-IR solar irradiation is determined in Section~\ref{solar:dimensions}. Then the effect of the oxide shell and particle dimensions on the optical properties is presented in Section~\ref{solar:oxide}. In Section~\ref{bb}, the optical response is then compared for layers under irradiation by a blackbody emitter at 1180~K. 

\section{Methods} \label{methods}
\subsection{Lorenz-Mie scattering of core@shell particles} \label{miemethods}
\begin{figure}[hbt!]
\includegraphics[width=\linewidth]{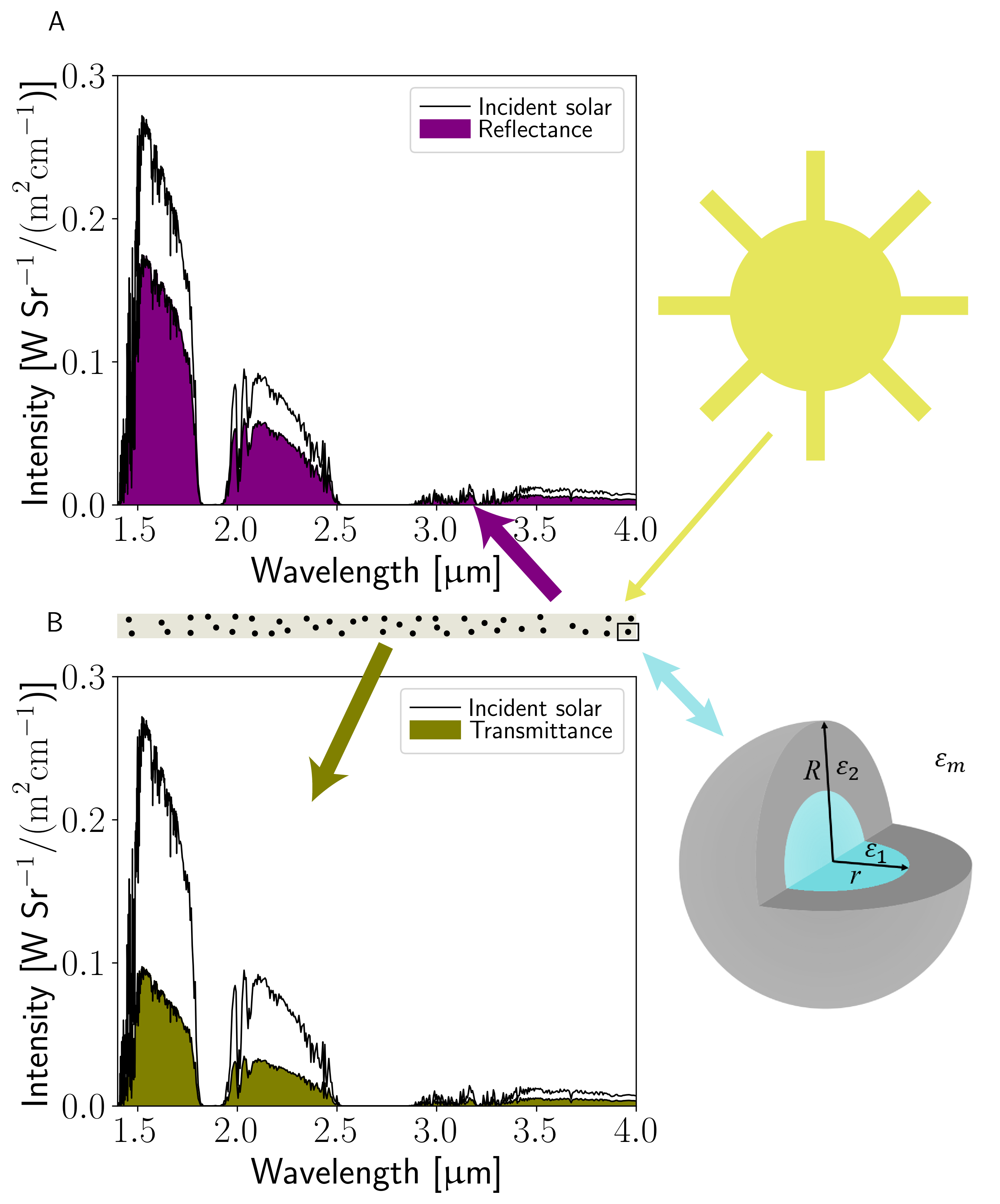}
\caption{\textbf{Solar Spectrum Schematic.} Calculated near-IR solar A) reflectance (purple area) and B) transmittance (green area) spectra of a 200~$\upmu$m thick layer embedded with Si@TiO$_2$ particles ($R$ = 0.6~$\upmu$m and $t$ = 0.3~$\upmu$m). The thickness of the shell, $t$, is the radius of the outer sphere, $R$, minus the radius of the inner sphere, $r$. For comparison, the near-IR solar spectrum is shown as black lines. The refractive index of medium is 1.5 and the particle volume fraction is 1\%.}
\label{fig:eff_cartoon}
\end{figure}

The dielectric response of small semiconductor@oxide (core@shell) spheres in an incident electromagnetic field is calculated using Lorenz-Mie theory.~\cite{aden1951scattering,bohren2008absorption} The core@shell spheres have a total radius $R$, core radius $r$, and shell thickness $t$ = $R$ -- $r$ (Figure~\ref{fig:eff_cartoon}). The oxide filling ratio, $\rho_{\rm{oxide}}$, is defined as the volume fraction of the oxide within the particle, \textit{i.e.} $\rho_{\rm{oxide}} = 1 - (r/R)^3$. Unless otherwise mentioned, the effect of the coating thickness was considered by replacing the surface layer with a homogeneous coating with varying thickness, $t$, for fixed total particle radius, $R$. 

The spherical particles, containing Si, InP, TiO$_2$, SiO$_2$, or ZrO$_2$, are surrounded by a non-absorbing insulating medium with constant refractive index ($n_{\rm{m}}$) of 1.5 and irradiated by near-IR light, $\lambda$ = 1.4 to 10~$\upmu$m. The materials were chosen from semiconductors with promising near-IR reflectance characteristics~\cite{tang2017plasmonically} and to cover a range of high- and low-$\kappa$ oxides. The bulk complex indices of refraction were obtained from Palik~\cite{palik1998handbook} or, in the case of ZrO$_2$, Wood and Nassau~\cite{wood1982refractive} and are shown in Figure S1. The bulk refractive indices of birefringent TiO$_2$ are averaged over the ordinary and the extraordinary directions. To represent IR grade fused silica, the extinction coefficient of silica is assumed to be zero below 3.5~$\upmu$m. The extinction coefficient of ZrO$_2$ was assumed to be zero and the refractive index restricted to wavelengths shorter than 5.1~$\upmu$m. 

The Lorenz-Mie coefficients, $a_n$ and $b_n$, were calculated using a program adapted from Bohren and Huffman,~\cite{bohren2008absorption} 
\begin{widetext}
\begin{equation}\label{eqn:an}
a_n = \frac{\psi_n(y)[\psi'_n(m_2y)-A_n\chi'_n(m_2y)]-m_2\psi'_n(y)[\psi_n(m_2y)-A_n\chi_n(m_2y)]}{\xi_n(y)[\psi'_n(m_2y)-A_n\chi'_n(m_2y)]-m_2\xi'_n(y)[\psi_n(m_2y)-A_n\chi_n(m_2y)]}\;,
\end{equation}
\begin{equation}
A_n = \frac{m_2\psi_n(m_2 x)\psi'_n(m_1 x) - m_1\psi'_n(m_2 x)\psi_n(m_1 x)}{m_2\chi_n(m_2 x)\psi'_n(m_1 x) - m_1 \chi'_n(m_2 x) \psi_n(m_1 x)}\;,
\end{equation}
\begin{equation}\label{eqn:bn}
b_n = \frac{m_2 \psi_n(y)[\psi'_n(m_2y)-B_n\chi'_n(m_2y)]-\psi'_n(y)[\psi_n(m_2y)-B_n\chi_n(m_2y)]}{m_2\xi_n(y)[\psi'_n(m_2y)-B_n\chi'_n(m_2y)]-\xi'_n(y)[\psi_n(m_2y)-B_n\chi_n(m_2y)]}\;,
\end{equation}
\noindent and 
\begin{equation}
B_n = \frac{m_2\psi_n(m_1 x)\psi'_n(m_2 x) - m_1\psi_n(m_2 x)\psi'_n(m_1 x)}{m_2\chi'_n(m_2 x)\psi_n(m_1 x) - m_1 \psi'_n(m_1 x) \chi_n(m_2 x)}\;
\end{equation}
\end{widetext}
\noindent where $m_1$ and $m_2$ are the refractive indices of the core and shell relative to the surrounding medium and the size parameters are $x = 2\pi r n_{\rm{m}}/\lambda$ and $y = 2\pi R n_{\rm{m}}/\lambda$. $\xi_n(z) = \psi_n(z)+ i \eta_n(z)$ are the Hankel functions of order $n$ wherein $\psi_n$ and $\eta_n$ are the spherical Bessel functions of the first and second kind, respectively. The Riccati-Bessel function $\chi_n(z)$ is $-zy_n(z)$ where $y_n(z)$ is a spherical Bessel function. Some of the limits to these equations are useful to mention. In the limit of zero core radius, lim$_{r\rightarrow 0} ~A_n$ = lim$_{r\rightarrow 0} ~B_n = 0$ and $a_n$ and $b_n$ reduce to those for a homogeneous sphere.  Also when the optical constants of the shell match the medium, i.e. $m_2 = 1$, the coefficients are equivalent to a sphere with radius $R$.

The far-field characteristics, that is the single particle efficiencies of scattering ($Q_{\textrm{sca}}$), absorption ($Q_{\textrm{abs}}$), extinction ($Q_{\textrm{ext}}$), and backscattering ($Q_{\textrm{back}}$) were calculated as
\begin{equation}\label{eqn:qsca}
Q_{\textrm{sca}} = \frac{2}{y^2} \sum_{n=1}^N (2n+1)(\lvert a_n \lvert^2 + \lvert b_n \lvert^2)\;,
\end{equation}
\begin{equation}
Q_{\textrm{abs}} = \frac{2}{y^2} \sum_{n=1}^N (2n+1)[\textrm{Re}(a_n + b_n) - (\lvert a_n \lvert^2 + \lvert b_n \lvert^2)]\;,
\end{equation}
\begin{equation}
Q_{\textrm{ext}} = Q_{\textrm{sca}} + Q_{\textrm{abs}}\;,
\end{equation}
\begin{equation}
Q_{\textrm{back}} = \frac{1}{y^2} \lvert \sum_{n=1}^N (2n+1)(-1)^n (a_n - b_n) \lvert^2
\end{equation}
\noindent where the summations are truncated at $N > y+4y^{1/3}+2$. The particle scattering asymmetry factor, $g$, is given by~\cite{bohren2008absorption}
\begin{eqnarray}
g = \frac{4}{y^2} \sum_{n=1}^N &&\big[ \frac{n(n+2)}{n+1} \textrm{Re}(a_n a^*_{n+1} + b_n b^*_{n+1} ) \nonumber \\
&&+ \frac{2n+1}{n(n+1)} \textrm{Re}(a_n b_n^*) \big]\;.
\end{eqnarray}

The possible values of the scattering asymmetry factor are between -1 and 1, and positive values indicate the scattered field is directed predominately forward while negative values indicate backscattering predominates.

\subsection{Optical properties of embedded layers} \label{mcmlmethods}
The spherical particles uniform in size and with a total radius varying from 0.1 to 4.0~$\upmu$m were considered to be randomly dispersed in a non-absorbing insulating matrix with constant refractive index 1.5. The compact layer thickness, $T$, was 200~$\upmu$m, such that $T >> R$. The compact layer was surrounded by a non-absorbing ambient medium. Unless otherwise specified the volume fraction of particles, $f$, was 1\%. 

The compact layers embedded with randomly dispersed core@shell spherical particles at low volume fraction are approximated with an effective medium such that the scattering and absorption coefficients, $\mu_{\textrm{sca}}$ and $\mu_{\textrm{abs}}$, of the compact layer are~\cite{tang2017plasmonically}
\begin{equation}
\mu_{\textrm{sca,abs}} = \frac{3}{2} \frac{fQ_{\textrm{sca,abs}}}{2R}\;,
\end{equation} 
\noindent and its effective dielectric permittivity, $\epsilon_{\textrm{layer}}$, from Maxwell Garnett Effective Medium Theory is~\cite{garnett1904}
\begin{equation}
\frac{\epsilon_{\textrm{layer}}-\epsilon_\textrm{m}}{\epsilon_{\textrm{layer}}+2\epsilon_\textrm{m}} = f_1\frac{\epsilon_1 - \epsilon_\textrm{m}}{\epsilon_1 + 2\epsilon_\textrm{m}} + f_2\frac{\epsilon_2 - \epsilon_\textrm{m}}{\epsilon_2 + 2\epsilon_\textrm{m}}
\end{equation}
\noindent where $\epsilon_1$, $\epsilon_2$, and $\epsilon_m$ are the dielectric permittivities of the core, shell, and medium components, and $f_i$ their constituent volume fractions. Of note is that $\epsilon_{\textrm{layer}}$ reduces to the effective medium of layer embedded with uncoated spheres when $f_1$ = 0, $f_2$ = 0, $\epsilon_1$ = $\epsilon_2$, or $\epsilon_2$ = $\epsilon_\textrm{m}$. 

The transmittance, reflectance ($\mathbb{R}$), and absorption of the free-standing compact layer were calculated using a modified Monte Carlo method~\cite{tang2017plasmonically} adapted from Wang et al.~\cite{wang1995mcml} The Monte Carlo method records the path and termination result of 10$^7$ photons from an infinitesimally small beam normal to the compact layer surface. The layer is described as an effective medium of the scattering and absorption coefficients, scattering anisotropy, and refractive index. Considering the cylindrical symmetry of the photon propagation, the grid resolution was $dz$ = 0.1~$\upmu$m and $dr$ = 5~$\upmu$m for the axial and radial directions, respectively. The diffuse reflectance and transmittance go to zero as a function of the radius of the layer. For the effective medium approximation to be valid, the near-field scattered radiation of one particle must not interact with another, and this is valid at low volume fraction, such as 1-6\% considered within this manuscript.  The core@shell method was verified against coated nanoparticles~\cite{laaksonen2013influence} and bare particles in the small-core limit.~\cite{tang2017plasmonically}

The reflected, absorbed, and transmitted photons are normalized to the number of photons. The reflectance of the layer under irradiation by an arbitrary incident source of photons normal to the compact layer surface is
\begin{equation}\label{eqn:reflectance}
    I^\textrm{i}_{\mathbb{R}}(\lambda) = \mathbb{R}(\lambda) I^\textrm{i}(\lambda)\;,
\end{equation}
where $I^\textrm{i}$ and $I^\textrm{i}_{\mathbb{R}}$ are the spectral densities of the incident light and total reflectance of the layer. The spectral densities of the transmittance and absorption of the layers are calculated in a similar manner. Two sources of near-IR radiation are considered. $I^\textrm{solar}$ is the spectral density of solar radiation standard AM 1.5G~\cite{standard1998g159} and $I^\textrm{BB}$ is the spectral density 23~m from the center of a blackbody emitter at 1180~K.~\cite{bordbar2019numerical} The solar and blackbody reflectance efficiency factors are 
\begin{equation}
\eta_{\textrm{solar,BB}} = \frac{\int_{\lambda_0}^{\lambda_1} \mathbb{R}(\lambda)I^{\textrm{solar,BB}}(\lambda) d\lambda}{\int_{\lambda_0}^{\lambda_1} I^{\textrm{solar,BB}}(\lambda) d\lambda}\;.
\end{equation}

\section{Results and Discussion} \label{results}
In this section, the near-IR optical properties of compact layers embedded with semiconductor particles are calculated using Lorenz-Mie theory and the Monte Carlo method described in Section~\ref{methods}. Each particle is a core@shell sphere (containing Si, InP, TiO$_2$, SiO$_2$, or ZrO$_2$) embedded in an insulating medium at low volume fraction (1\%). Particular attention is paid to the characterization of layers under irradiation by near-IR solar light in Section~\ref{solar} due to the availability of the AM 1.5G standard.~\cite{standard1998g159} In Section~\ref{bb} the optical response of the compact layers is analyzed upon irradiation by a blackbody emitter at 1180~K. 

\subsection{Optical response under solar irradiation} \label{solar}
The optical behavior of compact layers containing embedded semiconductor@oxide particles under near-IR solar radiation was calculated using a Monte Carlo method. Diffuse scattering by the embedded particles produces strong reflectance of near-IR radiation. Below, the scattering direction and energy is examined according to the particle dimension, material, and dielectric environment. The optical response is clarified according to the volume fraction and number density of particles within the layers. 

\subsubsection{Effect of particle dimensions} \label{solar:dimensions}
The optical response under solar irradiation is analyzed according to dimensions of particles embedded within a 200~$\upmu$m thick layer at a volume fraction of 1\%. The core radius, $r$, and total radius, $R$, of the spherical core@shell particles both affect the scattering behavior. At a fixed volume fraction, varying the total radius changes the particle number density. In this subsection, the oxide filling ratio, $\rho_{\rm{oxide}}$, is varied for particles with a fixed total size. Then, the results for particles with a different total size are compared. 

\begin{figure*}[hbt!]
\includegraphics[width=\linewidth]{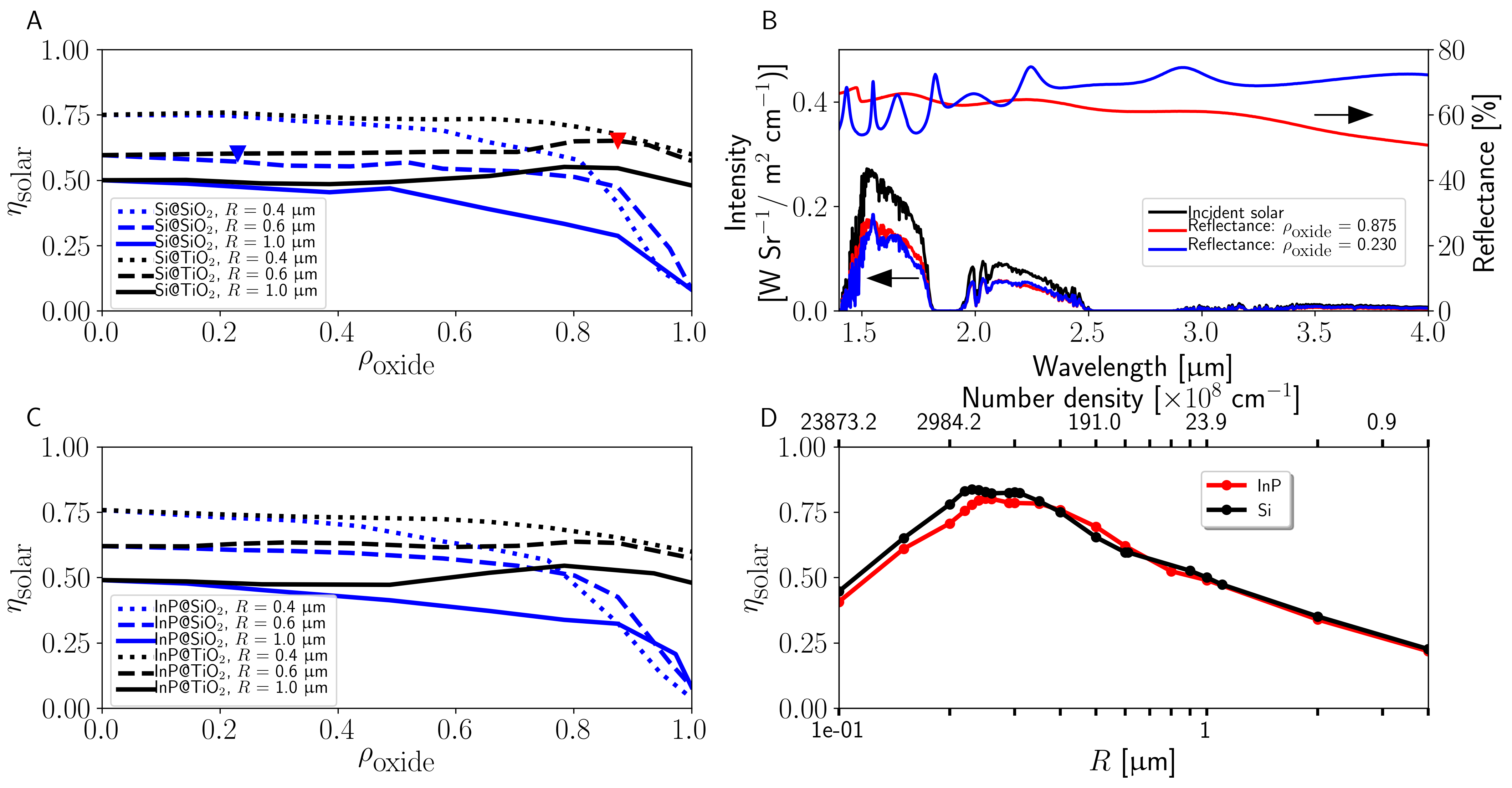}
\caption{\textbf{Solar Reflectance Efficiency Scan.} A, C) Solar efficiency of compact layers embedded with Si or InP particles coated with SiO$_2$ or TiO$_2$ with variable oxide filling ratio, $\rho_{\rm{oxide}}$ and total radius of 0.4, 0.6, or 1.0~$\upmu$m. B) Calculated total and solar reflectance spectra of layers containing Si@TiO$_2$ having a shell thickness of 0.05~$\upmu$m (blue lines) or 0.3~$\upmu$m (red lines) and total radius of 0.6~$\upmu$m. These particle dimensions correspond to $\rho_{\rm{oxide}}$ equal to 0.875 and 0.230, respectively, and are indicated by triangles in panel A. The black line is the incident near-IR solar spectrum. D) Calculated solar efficiency of layers containing bare Si or InP particles having a radius varying from 0.1 to 4.0~$\upmu$m. The maximum efficiency for layers containing bare particles is obtained for Si or InP particles with a radius of 0.23 or 0.25~$\upmu$m, respectively. The volume fraction is fixed at 1\% and the layer thickness is 200~$\upmu$m.} 
\label{fig:eff_scan}
\end{figure*}

The solar near-IR reflectance and transmittance spectra of a 200~$\upmu$m thick layer embedded at 1\% volume fraction with Si@TiO$_2$ oxide-coated semiconductor particles with $R$ = 0.6~$\upmu$m and $t$ = 0.3~$\upmu$m is provided in Figure~\ref{fig:eff_cartoon} and has been calculated using Eqn.~(\ref{eqn:reflectance}). For comparison, the near-IR spectral density of solar radiation standard AM 1.5G~\cite{standard1998g159} is shown as black lines. Despite the low volume fraction of embedded particles, the layers are efficient back reflectors and 65\% of the incident radiation is reflected. The photon direction within the low density embedded layer is quickly dispersed by the diffuse scattering of the particles. The scattering asymmetry of each core@shell particle is moderately forward scattering (Figures S2 and S3). The compact layer with embedded particles improves upon the reflectance of a layer without embedded particles that is characterized by specular reflectance alone. The specular reflectance obtained from our Monte Carlo simulations is approximately 4\% in all compacts considered.

The core size and shell thickness of the embedded particles affect the intensity and wavelength dependence of the reflected near-IR solar light. To illustrate the effects, the oxide filling ratio, $\rho_{\rm{oxide}}$, is varied from 0 to 1 for particles with a fixed total particle size. At $\rho_{\rm{oxide}}$ = 0 or 1, the solar reflectance efficiency, $\eta_{\rm{solar}}$, is equivalent to that of a bare semiconductor or oxide sphere, respectively. Sweeping $\rho_{\rm{oxide}}$ leads to local maxima and minima in the reflectance efficiency as the scattering modes align with the discrete peaks in the incident near-IR solar spectrum. The solar reflectance efficiencies of Si and InP particles coated with SiO$_2$ or TiO$_2$ with $R$ = 0.4, 0.6, and 1.0~$\upmu$m are shown in Figures~\ref{fig:eff_scan}A, C and those coated with ZrO$_2$ are given in Figure S4. 

The maximum solar reflectance efficiency of a layer embedded with Si@TiO$_2$ particles with $R$ = 0.6~$\upmu$m is 65\% at $\rho_{\rm{oxide}}$ = 0.875 (Figure~\ref{fig:eff_scan}A, black dashed line). This filling ratio corresponds to a shell thickness of 0.3~$\upmu$m and is marked by a red triangle. The total reflectance is intense and broad across the near-IR as seen in Figures~\ref{fig:eff_cartoon} and~\ref{fig:eff_scan}B (red curves). 

In layers embedded with particles having a different shell thickness, the reflectance is no longer as broad and intense which reduces the efficiency factor. The solar reflectance efficiency decreases to 60\% and 57\% in layers embedded with particles with $t$ = 0.05~$\upmu$m ($\rho_{\rm{oxide}}$ = 0.230 and marked with a blue triangle) or uncoated TiO$_2$ particles, respectively. The total and solar reflectance of the layer containing Si@TiO$_2$ particles with $R$ = 0.6~$\upmu$m and $t$ = 0.05~$\upmu$m is provided in Figure~\ref{fig:eff_scan}B (blue curves). The total reflectance is not as strong between the wavelengths about 1.5 to 1.8~$\upmu$m. However, there are sharp peaks with locally high reflectance. For example, a sharp reflectance peak is visible at 1.55~$\upmu$m in both the calculated total and solar reflectance spectra. 

Layers containing particles with a larger total radius (for example, $R$ = 1.0~$\upmu$m) are less efficient at reflecting the near-IR solar spectrum. Larger particles occupy a greater proportion of the total volume fraction, and do not significantly increase the scattering efficiency of each particle. Thus the inclusion of larger particles in layers diminishes the overall reflectance by reducing the particle number density. Compact layers containing small particles (for example, $R$ = 0.4~$\upmu$m) have a higher number density and tend to be more efficient at reflecting the near-IR solar spectrum. 

\begin{figure*}
\includegraphics[width=\linewidth]{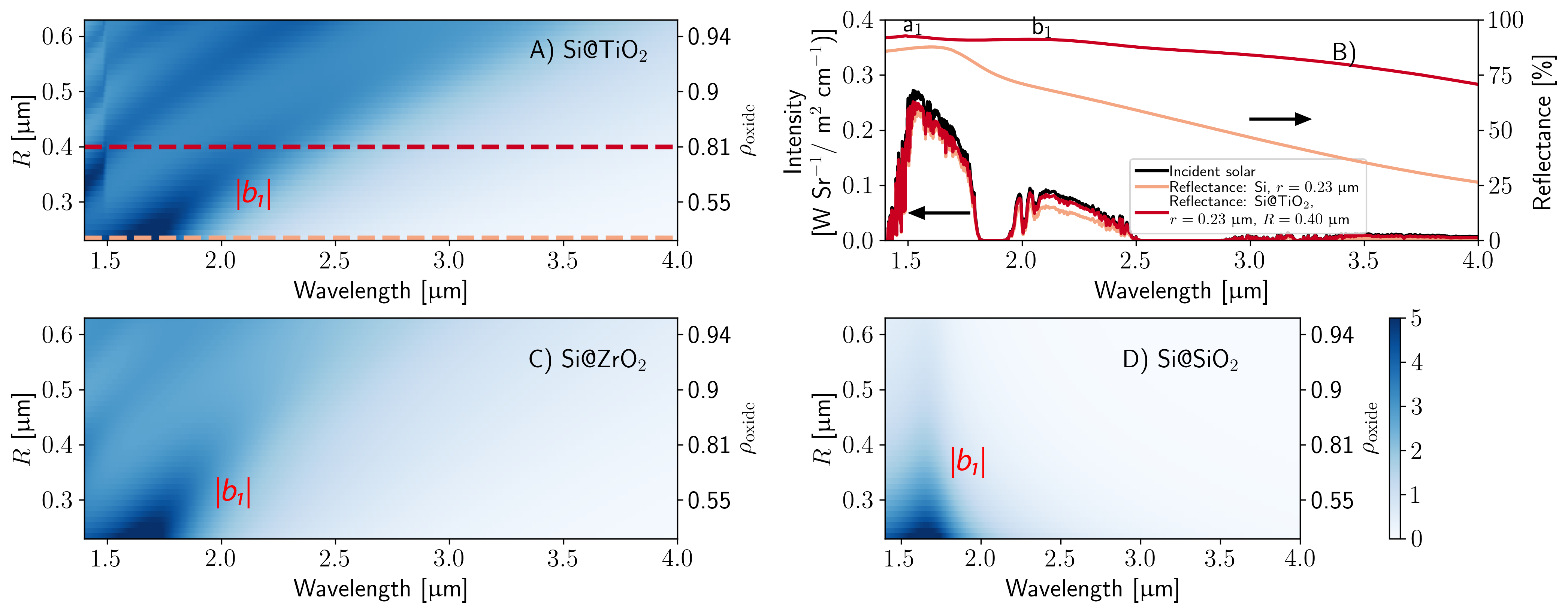} 
\caption{\textbf{Single particle scattering efficiency of Si@oxide particles.} $Q_{\textrm{sca}}$ of Si particles coated with (A) TiO$_2$, (C) ZrO$_2$, or (D) SiO$_2$ with fixed $r$ = 0.23~$\upmu$m. B) The calculated optical response of 200~$\upmu$m thick compact layers embedded with bare or coated particles with a shell thickness of 0.17~$\upmu$m ($\rho_{\rm{oxide}} = 0.810$). The particle number density is equivalent to a layer containing 1\% volume fraction of the bare particle (1960~$\times 10^8$ particles per~cm$^{3}$). The solar reflectance efficiency of the layer containing bare Si is 83.7\% and the layer containing particles coated with TiO$_2$ was 91.3\%.}
\label{fig:Qsize_scan}
\end{figure*}

The self-cannibalization is evident in the decline of the reflectance efficiency of layers embedded with large bare particles (Figure~\ref{fig:eff_scan}D). The sparsity of the larger particles in the layers is clear from the number density along the secondary $x$-axis. For very small particles, the reflectance efficiency decreases despite the increase of the number density, and is due to the low scattering efficiency of these particles at longer wavelength. The maximum solar reflectance efficiency is 83.7\% and 80.2\% for a total radius of 0.23~$\upmu$m and 0.25~$\upmu$m in 200~$\upmu$m thick layers containing 1\% Si or InP particles, respectively. The scattering efficiencies, $Q_{\textrm{sca}}$, of bare and coated Si and InP particles are shown in Figures S5--S7. 

\subsubsection{Effect of the oxide shell} \label{solar:oxide}
In this subsection, the energies, magnitude, and directional field of the scattering modes are discussed in terms of the material and thickness of the oxide coating. To clearly show the influence of shell thickness, the core size is fixed and the total particle size is varied. To compare the optical properties of compact layers embedded with particles of different total size, the number density is fixed and specified.

The scattering efficiency of Si@oxide single particles with a core radius fixed at 0.23~$\upmu$m is presented in Figure~\ref{fig:Qsize_scan}. The layer containing bare Si particles of this size had the highest solar reflectance efficiency in the previous sub-section. The shell thickness varies along the $y$-axis and the dark blue areas indicate the wavelengths where the magnitude of $Q_{\textrm{sca}}$ is large. The oxide coating dampens the magnitude of $Q_{\textrm{sca}}$ compared to bare particles,~\cite{laaksonen2013influence,seyedheydari2020near} and yet the optical behavior of the layers is influenced by factors such as the alignment of the scattering modes with the discrete near-IR solar spectrum and the direction and magnitude of scattered field. These factors are discussed below.

The addition of an oxide coating shifts the energies of the scattering modes from the semiconductor core by changing the dielectric environment. The scattering modes of the ZrO$_2$- and TiO$_2$-coated particles are redshifted and the extent of the shift is greater for thicker coatings. The magnitude of the shift is larger in the particles coated with TiO$_2$ than with either SiO$_2$ or ZrO$_2$. This material effect is expected due to the value of $m_2$, the relative refractive index of oxide shell and the medium, which is greater than one for ZrO$_2$- and TiO$_2$.~\cite{laaksonen2013influence,seyedheydari2020near}

The shifting of the scattering modes affects the alignment with the incident solar spectrum. There are minor fluctuations in the maximum reflectance efficiencies between the different oxides in Figure~\ref{fig:eff_scan}A, C, but the effect is more clearly observed in the single particle scattering efficiency, $Q_{\textrm{sca}}$, with a fixed core particle radius. To illustrate the improvement of the solar reflectance efficiency the bare Si particle is compared to a Si@TiO$_2$ particle with a shell thickness of 0.17~$\upmu$m ($\rho_{\rm{oxide}} = 0.810$). The scattering efficiency of each particle is indicated by the horizontal lines in Figure~\ref{fig:Qsize_scan}A. The application of the shell shifts the lowest energy mode, the dipolar magnetic mode, $b_1$, from approximately 1.6~$\upmu$m to 2.0~$\upmu$m. The mode was identified by direct inspection of the Mie coefficients (Figure S8). The energies of the dipolar magnetic mode and dipolar electric mode align better with the incident solar spectrum as seen in the calculated total and solar reflectance in Figure~\ref{fig:Qsize_scan}B. The TiO$_2$ coating increases the solar reflectance efficiency of the layer from 83.7\% to 91.3\% for a fixed number density (1960~$\times 10^8$ particles per~cm$^{3}$). By comparison, the layers containing SiO$_2$- and ZrO$_2$-coated particles have solar reflectance efficiency factors of only 83.0\% and 89.0\%, respectively. Of the three layers with oxide-coated particles, only the layer with SiO$_2$-coated particles is less reflective than the one containing bare particles. By comparing the layers based on equivalent particle number density but different total size, the particle volume fraction of the layer containing Si@oxide particles has increased to 5.3\%.

The scattering by the semiconductor@oxide particles is more efficient for oxides having larger relative refractive indices of the shell and medium.~\cite{laaksonen2013influence,seyedheydari2020near} The magnitude of the scattering is exhibited in the higher reflectance efficiency of layers containing ZrO$_2$- or TiO$_2$-coated particles than layers embedded with SiO$_2$-coated particles (Figures~\ref{fig:eff_scan}A, C and S4). The effect is more pronounced at large oxide filling fraction. In addition, the magnitude of the scattering efficiency of the Si@oxide particles with fixed core size can be directly compared in Figure~\ref{fig:Qsize_scan}. 

\begin{figure}
\includegraphics[width=\linewidth]{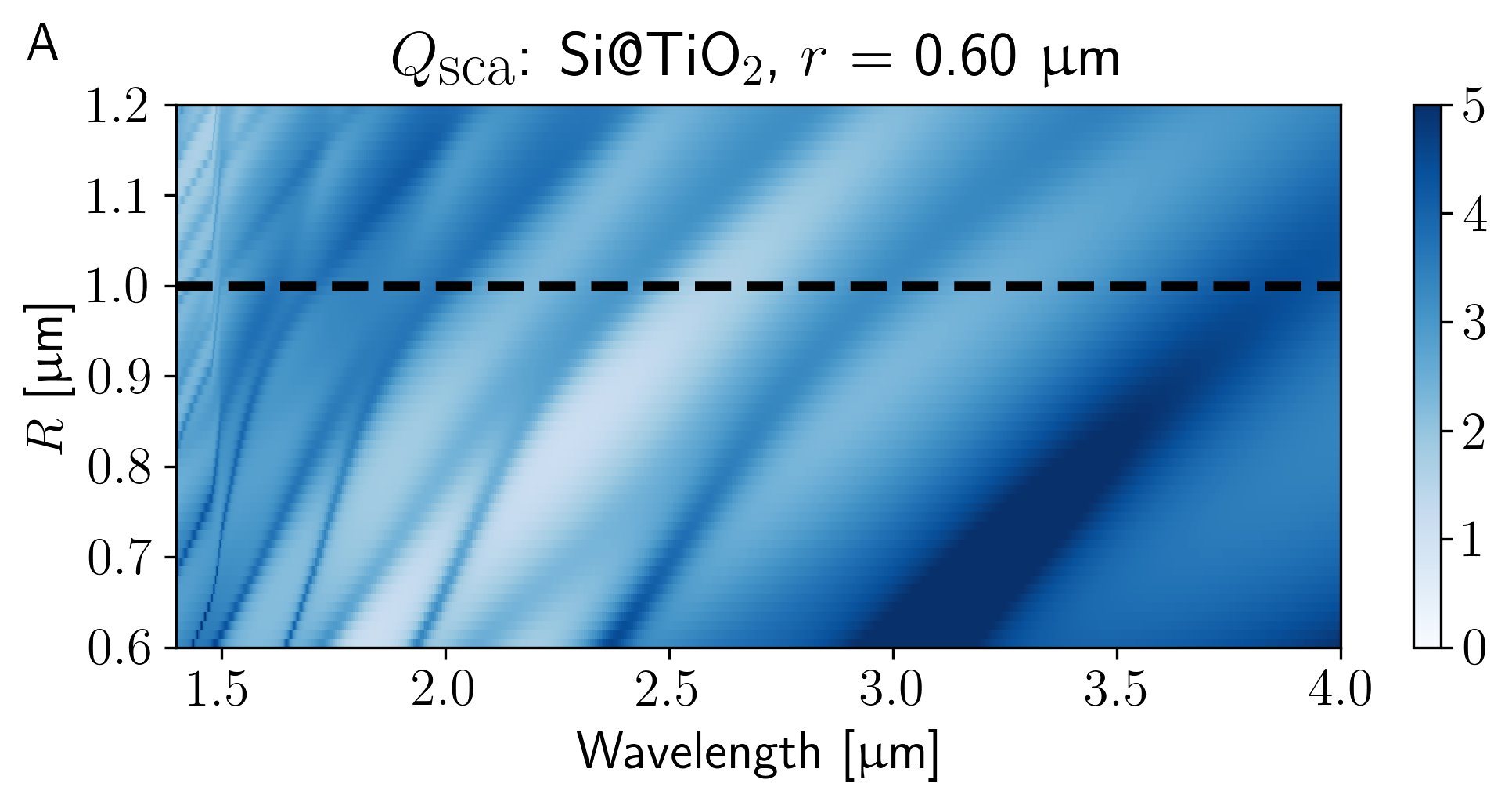}\\
\includegraphics[width=\linewidth]{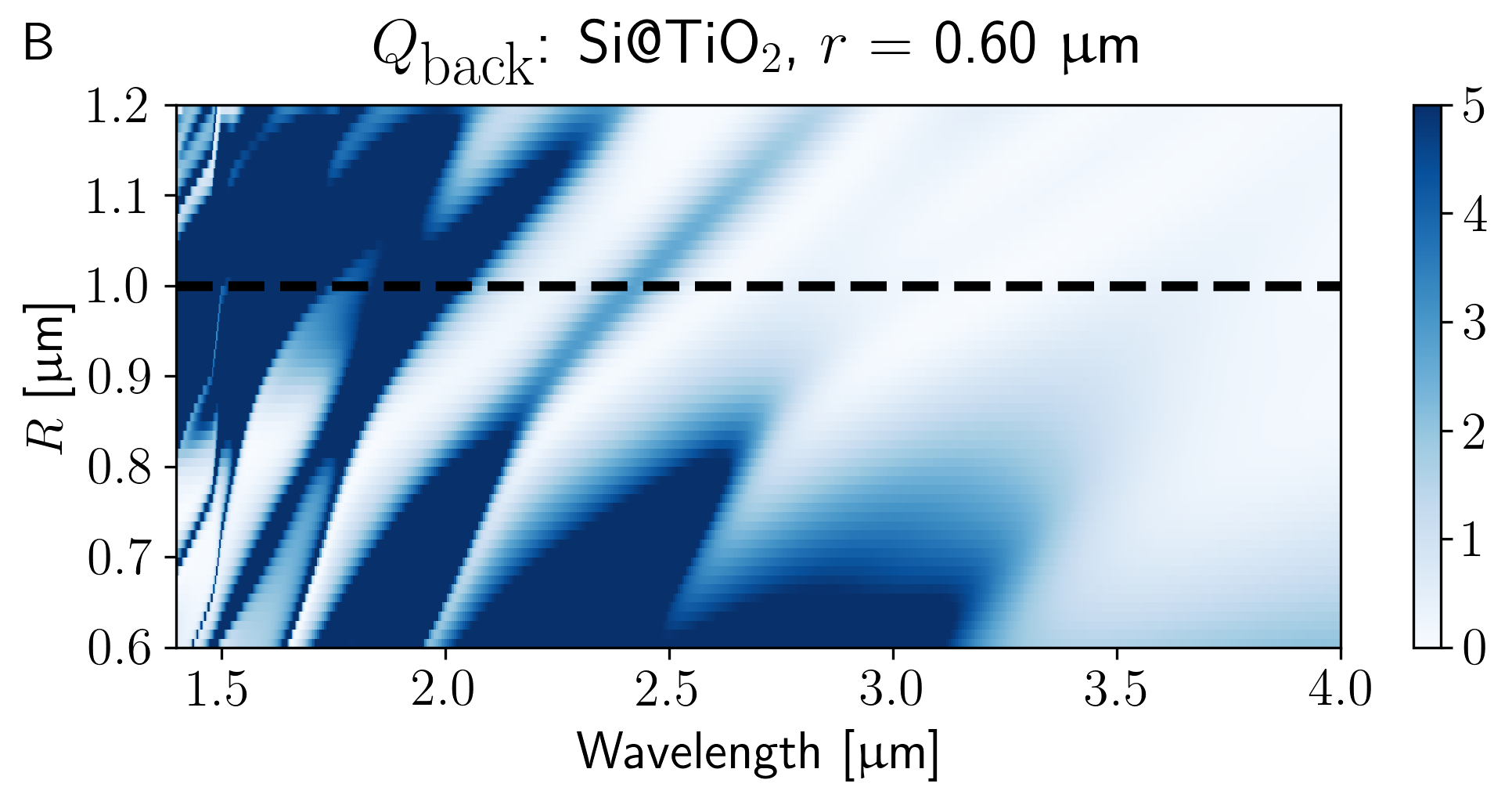}
\caption{\textbf{Single particle scattering efficiency of Si@TiO$_2$.} (A) Total scattering efficiency, $Q_{\textrm{sca}}$, and (B) backscattering efficiency, $Q_{\textrm{back}}$, of Si particles coated with TiO$_2$ and fixed core radius, $r$, equal to 0.60~$\upmu$m. The dashed lines indicate the total and backscattering efficiencies of the particle with a 0.4~$\upmu$m thick shell.}
\label{fig:qback}
\end{figure}

The oxide shell modifies the scattered field and this is demonstrated by the backscattering efficiency, $Q_{\textrm{back}}$ and total scattering efficiency, $Q_{\textrm{sca}}$, of the low energy modes in a Si@TiO$_2$ particle with fixed core radius of 0.6~$\upmu$m (Figure~\ref{fig:qback}). The composite nature of the higher energy modes ($\lambda <$~2.5~$\upmu$m) makes the behavior less clear. The magnitude of the backscattering efficiency for the low energy modes reduces substantially with growing shell thickness despite the total scattering efficiency remaining strong. For example, the addition of a 0.4~$\upmu$m thick shell (indicated by the dashed lines in Figure~\ref{fig:qback}) decreases the backscattering efficiency from 5.9 (at $\lambda$ = 2.58~$\upmu$m) in the bare particle to 0.4 (at $\lambda$ = 3.56~$\upmu$m) at the quadrupolar electric mode energy, $a_2$. By contrast, the total scattering efficiency remains strong (declining from 3.6 to 3.4). The sustained total scattering and reduced backscattering magnitude indicates the scattered field has been directed forward. This behavior is a result of the increased overlap of the quadrupolar electric mode and the quadrupolar magnetic mode which approximately satisfies a Kerker condition for minimum backscattering efficiency.~\cite{kerker1983electromagnetic,tzarouchis2018light} The modes and magnitudes were identified by direct inspection of the Mie coefficients and efficiencies, and the forward shift of the scattered field is validated by the increase of the scattering anisotropy factor (Figure S9). The suppression of the far-field backscattering was previously demonstrated in metal@dielectric (core@shell) particles and Si disks.~\cite{liu2012broadband,staude2013tailoring}

\begin{figure}[htbp!]
\includegraphics[width=\linewidth]{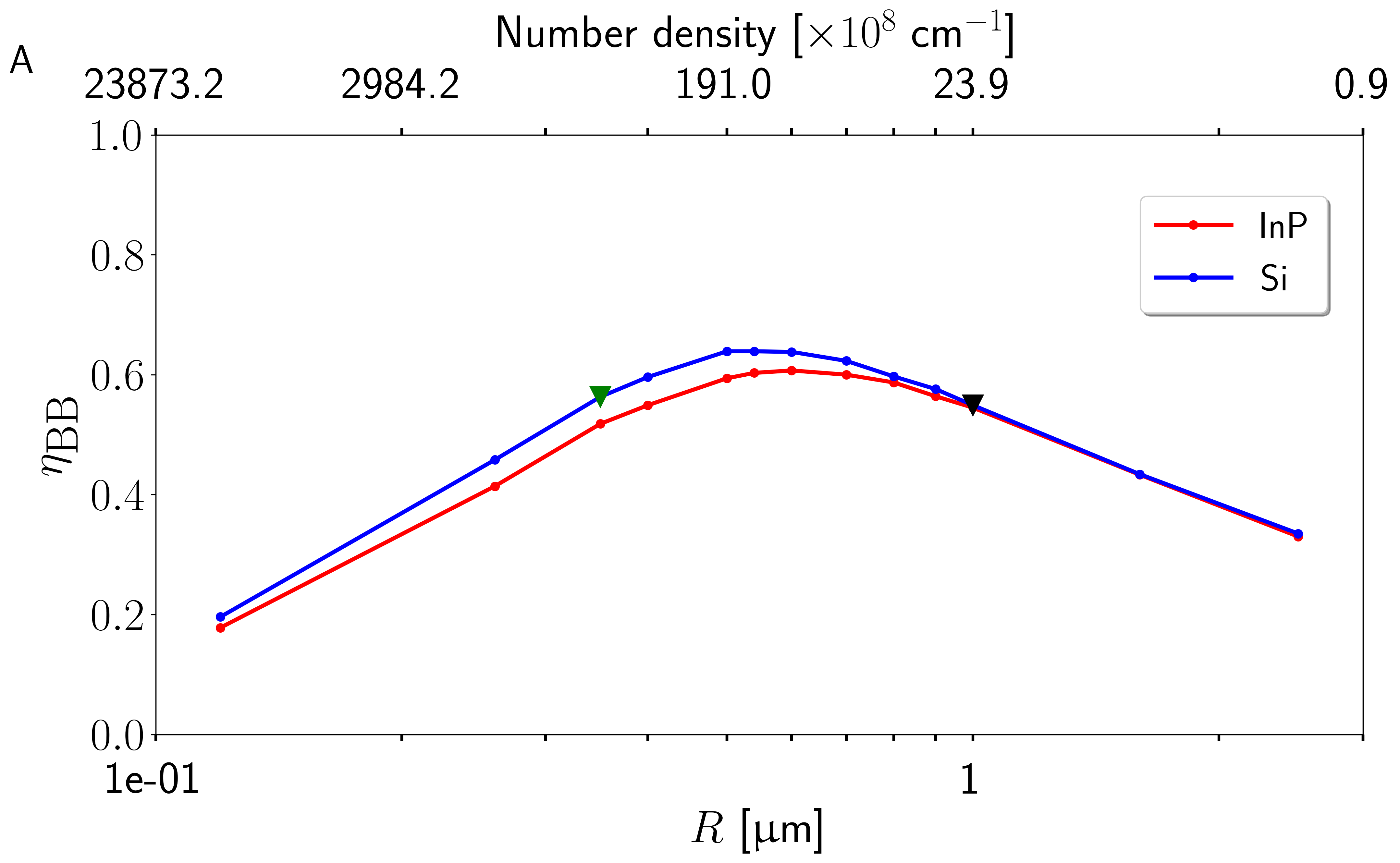}\\
\includegraphics[width=\linewidth]{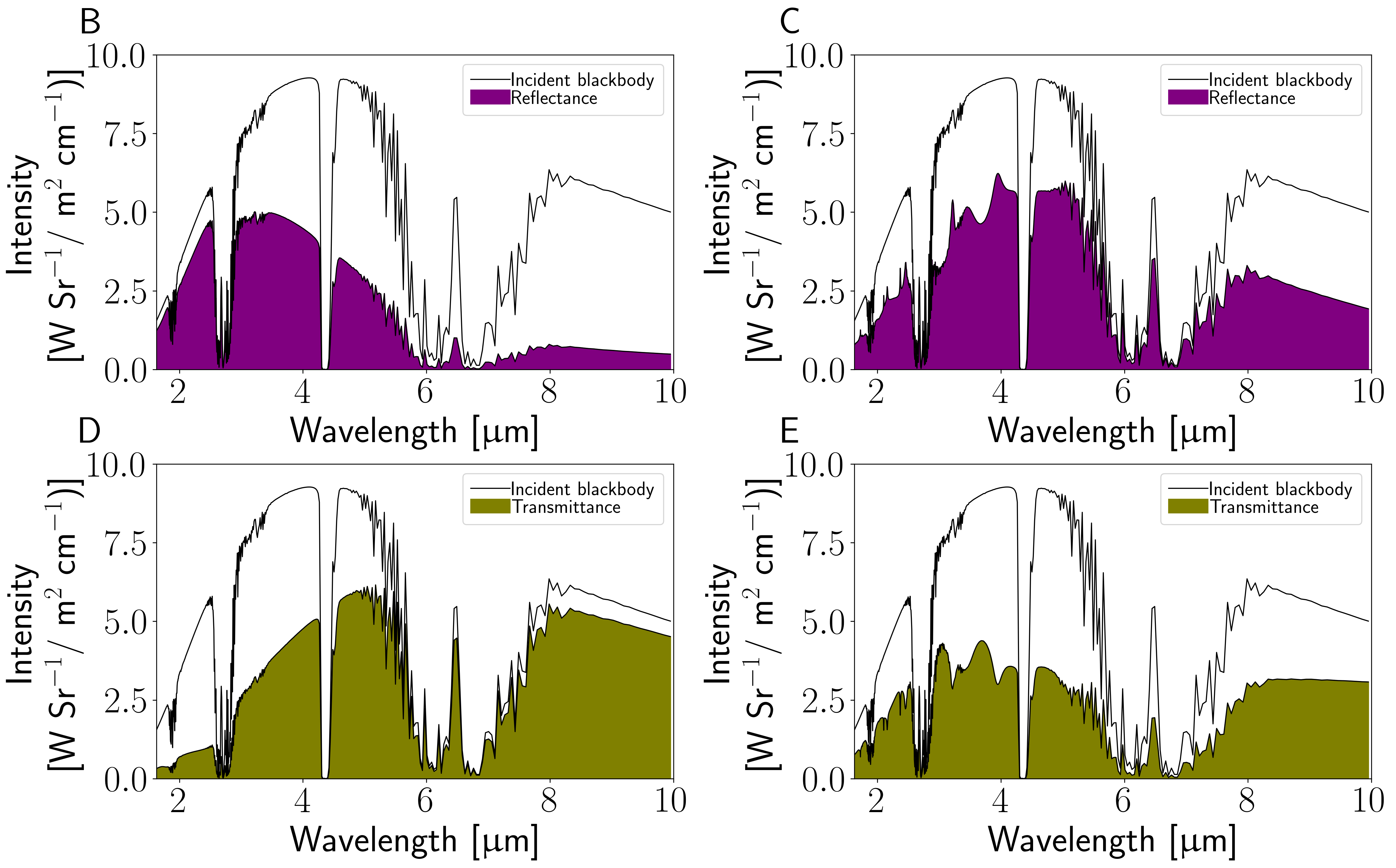} 
\caption{\textbf{Particle size and blackbody radiation.} A) Calculated blackbody reflectance efficiency of 200~$\upmu$m thick layers containing bare Si or InP particles having a radius varying from 0.12 to 2.5~$\upmu$m. B-E) Calculated reflectance and transmittance spectra of layers embedded with semiconductor particles and irradiated from a blackbody emitter at 1180~K at a distance of 23~m. The reflected or transmitted light is filled proportionally under the incident spectrum (black curves). The layers are embedded with Si particles having a radius of (B, D) 0.35~$\upmu$m or (C, E) 1.00~$\upmu$m. The blackbody reflectance efficiency of these layers are 56.3\% and 54.9\% and marked with green and black triangles in (A), respectively. The refractive index of medium is 1.5 and the particle volume fraction is 1\%.}
\label{fig:bb_scan}
\end{figure}

\begin{figure*}[hbt!]
\includegraphics[width=\linewidth]{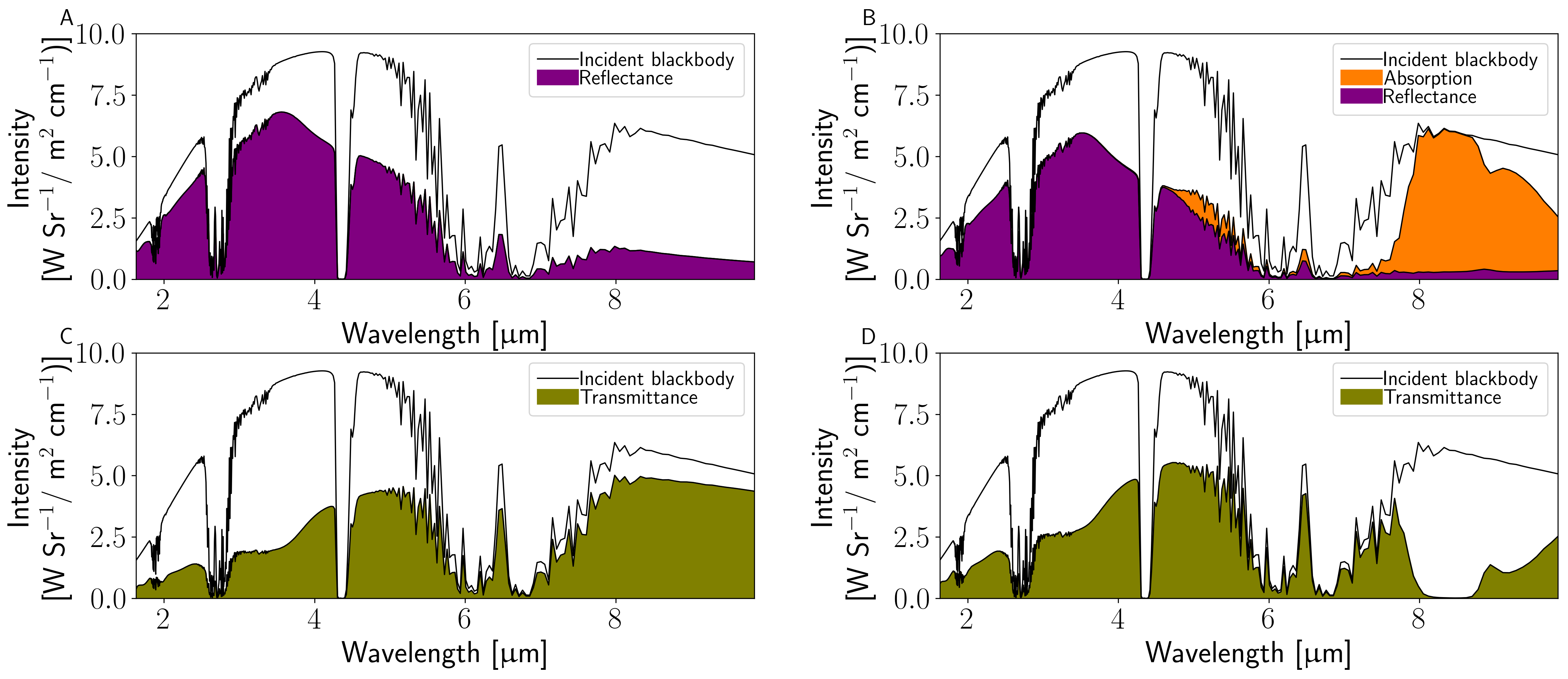}\\
\includegraphics[width=\linewidth]{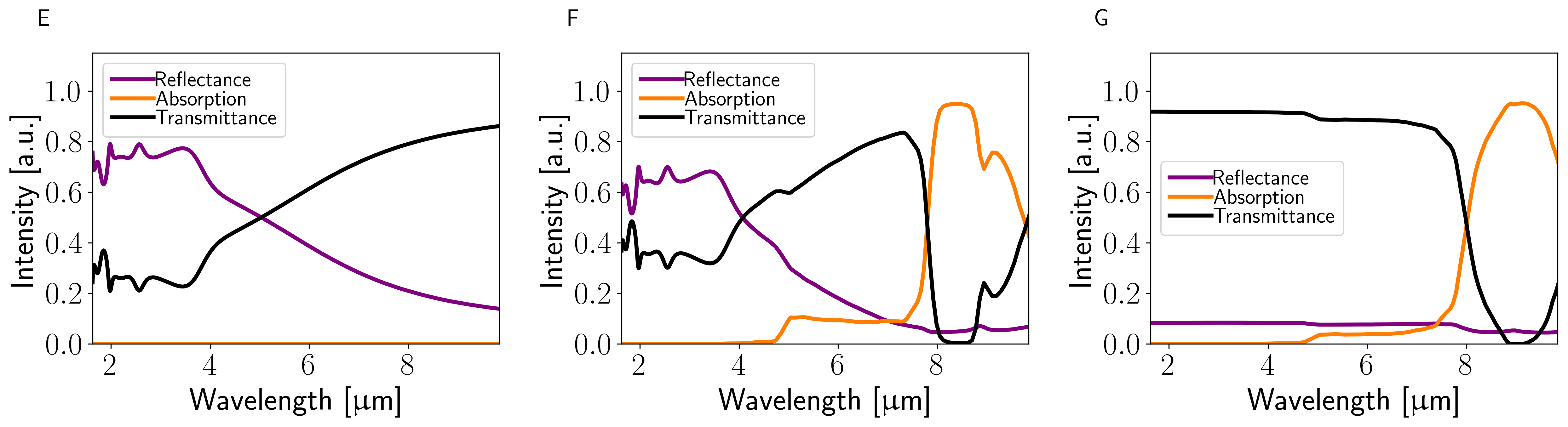}
\caption{\textbf{Response of compact layers to blackbody radiation.} A-D) Reflectance and transmittance spectra of a 200~$\upmu$m thick layers embedded with semiconductor particles and irradiated from a blackbody emitter at 1180~K. The reflected, absorbed, or transmitted light is filled proportionally under the incident spectrum (black curves). The embedded particles are (A, C) Si with a total radius of 0.50~$\upmu$m or (B, D) Si@SiO$_2$ with an core radius of 0.50~$\upmu$m and a total radius of 0.6~$\upmu$m.  Calculated optical properties of layers containing (E) bare Si, (F) Si@SiO$_2$, or (G) bare SiO$_2$ particles ($r$ = 0.50~$\upmu$m, $t$ = 0 or 0.1~$\upmu$m). The refractive index of medium is 1.5 and the particle volume fraction is 1\%.}
\label{fig:bb_cartoon}
\end{figure*}

The compact layers remain highly reflective despite containing particles with low backscattering efficiency. The solar reflectance efficiency is 55.1\% in the layer embedded with coated particles ($r$ = 0.6~$\upmu$m, $t$ = 0.4~$\upmu$m) at 1\% volume fraction. By comparison, the solar reflectance efficiency is 59.6\% in the layer embedded with bare particles. Thus the high reflectance is achieved using the diffuse scattering of particles. Notably, if the layers are compared at a number density equivalent to a layer containing 1\% volume fraction of bare particles having a radius of 0.6~$\upmu$m (110~$\times 10^8$ particles per~cm$^{3}$), the reflectance efficiency of the layer embedded with coated particles is 80.4\%. The enhanced reflectance emphasizes the diffuse nature of the scattering processes within the layer. 

\subsection{Optical response under irradiation from a blackbody emitter} \label{bb}
In this subsection, the optical behavior of semiconductor@oxide-embedded compact layers was calculated under irradiation from a blackbody emitter. The incident spectral density was obtained from previously published simulations of the spectral density at a distance of 23~m from the center of a blackbody emitter at 1180~K.~\cite{bordbar2019numerical} The numerical simulation used a line-by-line integration method to solve the spectral radiative transfer equation along a line of sight and included the effects of molecular absorption and emission.~\cite{bordbar2019numerical} The intervening medium contained atmospheric air at 20\degree C with a mole fraction of 0.0004 and 0.004 for CO$_2$ and H$_2$O, respectively.~\cite{bordbar2019numerical} The absorption of gases produces a discrete spectrum. The incident spectrum extends deeper into the IR region ($\lambda$ = 1.6 to 10~$\upmu$m) than the solar spectrum considered in the previous subsection. In addition to the particle size and material effects described in the preceding subsection, the absorption from the oxide shell must be considered.

\subsubsection{Effect of particle dimensions}
The calculated reflectance efficiency of 200~$\upmu$m thick compact layers containing bare Si or InP particles of different size and a volume fraction fixed at 1\% when irradiated by a blackbody emitter at 1180~K is shown in Figure~\ref{fig:bb_scan}A. The maximum reflectance efficiency of 63.9\% and 60.7\% is achieved in layers containing Si or InP particles with a radius of 0.50 or 0.60~$\upmu$m, respectively. The maximum blackbody reflectance efficiency occurs for larger particle sizes than under solar irradiation. Blackbody radiation extends deeper into the IR region than solar radiation, and the scattering by smaller particles is not sufficiently broadband to cover the full blackbody spectrum. Larger particles are needed to efficiently reflect the full blackbody range. Similar to the solar reflectance efficiency in Figure~\ref{fig:eff_scan}D, there is a trade-off between large particles and low number density. Compact layers containing very large particles are less efficient due to the reduced number density at fixed volume fraction. 

\begin{figure}[hbt!]
\includegraphics{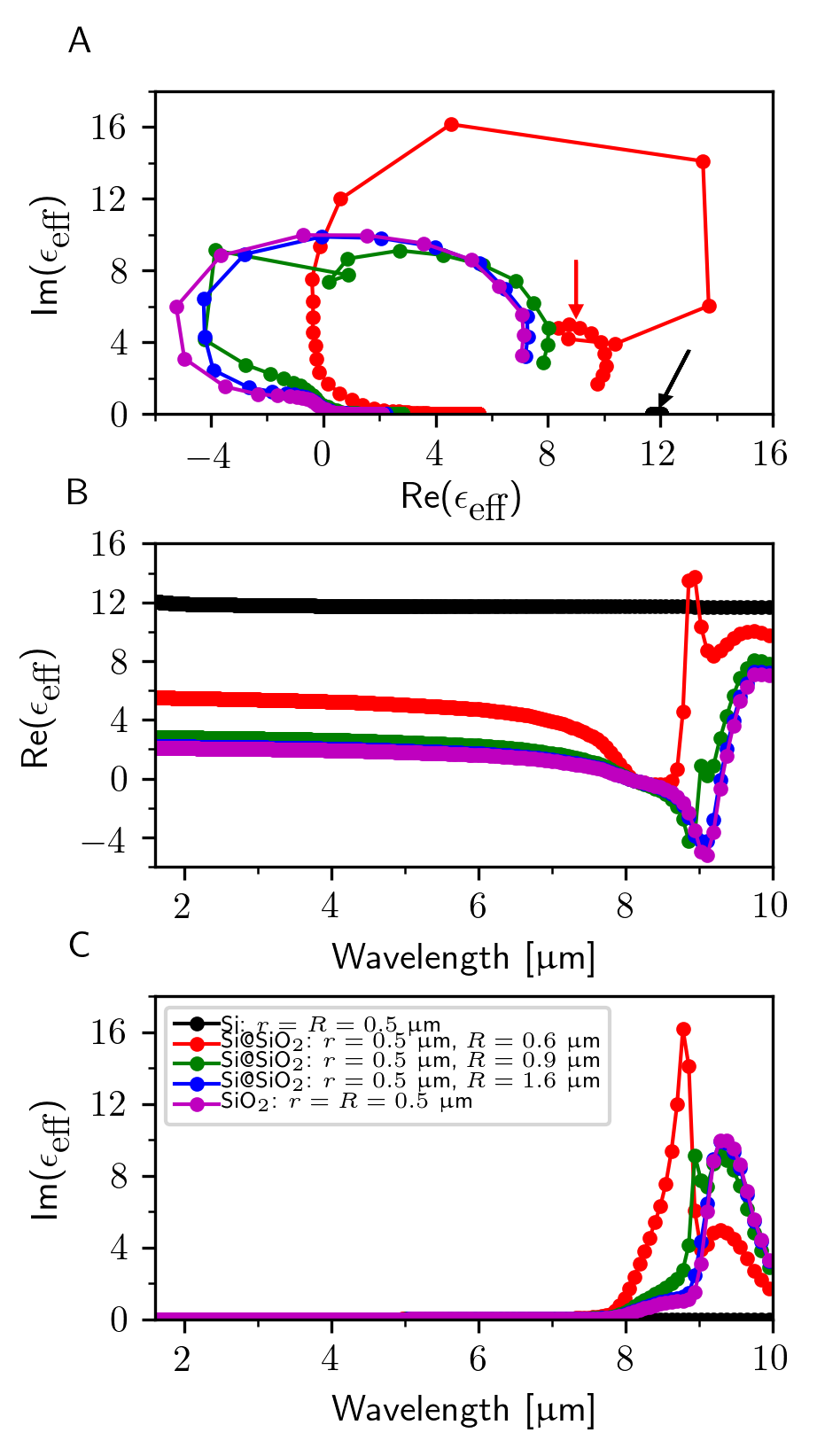}
\caption{The effective permittivity of Si@SiO$_2$ particles presented as A) a Cole-Cole plot and the B) real and C) imaginary parts of the permittivity as functions of the wavelength. The core radius is 0.5~$\upmu$m and the total radius varies from 0.5 to 1.6~$\upmu$m. The permittivity of SiO$_2$ is also shown for comparison.}
\label{fig:effective-per}
\end{figure}

The calculated reflectance of the layers containing bare Si particles with a radius of 0.35~$\upmu$m and 1.00~$\upmu$m are shown in Figure~\ref{fig:bb_scan}B-E. For comparison, the spectral density of the incident blackbody radiation is shown as black lines. The efficiencies are approximately the same (56.3\% and 54.9\%, respectively) and yet the reflectance spectra are distinct. The layer embedded with smaller particles strongly reflects the short wavelengths (2 to 4~$\upmu$m) and transmits the long wavelengths (7 to 10~$\upmu$m). The layers containing the larger particles have a broader reflectance across the spectrum. 

\subsubsection{Effect of the oxide shell}
The effect of the oxide shell on the shift of the scattering mode energies, magnitude, and direction is similar to the discussion for the solar spectrum in Section~\ref{solar:oxide}. In addition, the broad spectral range of the blackbody emitter means photoabsorption from the oxide must be considered. Absorption has been negligible in the layers considered thus far. In the range considered here, the absorption by the semiconductors is low (bandgap of Si and InP is 1.1 and 0.92~$\upmu$m, respectively), but oxides often have large extinction coefficients at longer wavelengths. For example, SiO$_2$ has Si-O stretching modes between about 7 -- 10~$\upmu$m.~\cite{morrow1992surface} Thus, when the compact layers containing particles coated with SiO$_2$ are under irradiation by a blackbody emitter, absorption competes with scattering in photon transport.

To illustrate the effect, the optical response under irradiation from the blackbody emitter was calculated for 200~$\upmu$m thick layers embedded with Si@SiO$_2$ particles with a core radius of 0.50~$\upmu$m and a shell thickness of 0 or 0.1~$\upmu$m and volume fraction of 1\% (Figure~\ref{fig:bb_cartoon}A-D). The corresponding reflectance, absorption, and transmittance spectra are presented in Figure~\ref{fig:bb_cartoon}E, F. The addition of a thin oxide coating leads to absorption at wavelengths longer than 4~$\upmu$m. The absorption within the layer reduces its broadband reflectance. 

The absorption band in the layer containing particles with thin SiO$_2$ coatings have more features than the band in the compact layer with bare SiO$_2$ particles (Figure~\ref{fig:bb_cartoon}G). The effective permittivity of the core@shell particles provides insight into the formation of these rich features and other effects of the oxide. The quasi-static approximation is used to obtain an effective permittivity of a core@shell sphere, $\epsilon_\textrm{eff}$, and is given by~\cite{chettiar2012internal}
\begin{equation}
\epsilon_\textrm{eff} = \epsilon_2 \frac{R^3 (\epsilon_1+2\epsilon_2) + 2r^3 (\epsilon_1-\epsilon_2) }{R^3 (\epsilon_1+2\epsilon_2) - r^3 (\epsilon_1-\epsilon_2)}\;.
\label{eqn:per-core@shell}
\end{equation}
The effective permittivity of Si@SiO$_2$ with various particle dimension is shown as Cole-Cole plots~\cite{cole1941dispersion,bohren2008absorption} and as a function of the wavelength in Figure~\ref{fig:effective-per}. The effective permittivity of the other core@shell particles are provided in Figures S10 and S11. The permittivity of the bare Si is relatively constant in this region and is indicated by a black arrow. The relatively constant behavior at a large positive value of the real part of the permittivity suggests the dielectric nature of the modes. Upon addition of a small amount of oxide, the permittivity and Cole-Cole plot develop more features. At shorter wavelengths (1 to 7~$\upmu$m), the real part of the permittivity decreases toward the permittivity of bulk SiO$_2$. In the Cole-Cole plot, the real part of the effective permittivity shifts to smaller values while the imaginary part remains zero. At longer wavelengths, the permittivity curls clockwise as the absorption bands change the permittivity. 

The core@shell particles with thin shells of SiO$_2$ have a more intricate dielectric permittivity than the bulk SiO$_2$. For example, a tight loop in the Cole-Cole plot of the particle with $R$ = 0.6~$\upmu$m (red curve) is indicated by the red arrow. The loop corresponds to a local minimum in the real part of the effective permittivity at $\lambda$ = 9.2~$\upmu$m. The rich features are manifested in the absorption band of the respective layers (Figure~\ref{fig:bb_cartoon}E-G). The absorption band in the layer containing particles with thin SiO$_2$ coatings have more features than the band in the layer with bare SiO$_2$ particles. 

\section{Conclusions} \label{conc}
The optical properties of 200~$\upmu$m thick compact layers containing semiconductor@oxide spherical particles with a total radius varying from 0.1 to 4.0~$\upmu$m at low volume fraction were calculated using Lorenz-Mie theory and a Monte Carlo method. Even at low volume fraction and layer thickness, the embedded core@shell spheres directionally scatter light which produces strong reflectance in the near-IR region. The optical response of the compact layers under irradiation by solar and blackbody sources was clarified according to the particle size and its relationship to the scattering efficiency and particle number density. 

Si and InP particles with a radius of 0.23~$\upmu$m and 0.25~$\upmu$m were the most effective at reflecting near-IR solar radiation ($\eta_\textrm{solar}$ = 83.7\% and 80.2\%, respectively) in 200~$\upmu$m thick layers containing embedded particles at 1\% volume fraction. Larger particles were needed to obtain the maximum reflectance efficiency of compact layers at a distance of 23~m from the center of a blackbody emitter at 1180~K due to the spectral range extending deeper into the IR. The blackbody reflectance efficiency was 63.9\% and 60.7\% for compact layers containing bare Si and InP particles with a radius of 0.50 or 0.60~$\upmu$m, respectively. 

The solar reflectance efficiency factor increased to over 90\% provided the alignment with the incident spectral density was improved. The addition of an oxide shell (SiO$_2$, ZrO$_2$, or TiO$_2$) shifted the scattering mode energies by modifying the dielectric environment of the core. The calculations demonstrate that inclusion of larger particles reduces the particle number density and does not significantly increase the scattering efficiency of each particle. 

The large scattering efficiencies and good thermal stability of the semiconductor particles could be exploited in devices to direct the propagation of near-IR radiation and limit unwanted radiative and conductive heat exchange in compact layers for thermal insulation~\cite{tang2017plasmonically} or solar cell applications. Particular focus in this manuscript was paid to the optical response of the compact layers under irradiation from solar and blackbody sources, but the compact layers are also responsive to other incident near-IR sources, such as flame emission. The selective reflectance or transmittance properties of the particle-embedded layers is expected to differentiate radiation sources in array-based sensing technologies. The applications benefit from the facile manufacture of low volume fraction compact layers. 

\section*{Supplementary Material}\label{SI}
See supplementary material for the bulk and effective permittivities of the materials and core@shell particles and for the scattering efficiencies and asymmetry factors.

\section*{Data Availability}\label{data}
The data that supports the findings of this study are available within the article and its supplementary material.

\begin{acknowledgments}
We acknowledge Dr. Hadi Bordbar, Aalto University, for providing the incident blackbody spectrum. This work was performed as part of the Academy of Finland project 314488 and QTF Centre of Excellence program (project 312298) (KC, FS, TAN). We acknowledge computational resources provided by CSC -- IT Center for Science (Finland) and by the Aalto Science-IT project (Aalto University School of Science); the Discovery Grants and Canada Research Chairs Program of the Natural Sciences and Engineering Research Council (NSERC) of Canada; and Compute Canada (www.computecanada.ca).
\end{acknowledgments}

%\nocite{*}
\bibliography{MCbib}

\end{document}